\documentclass [aps,twocolumn,superscriptaddress,altaffilletter,lengthcheck,tightenlines,showpacs,showkeys]{revtex4}
\usepackage[dvipdf]{epsfig}

\newcommand{\ben}{\begin{eqnarray}}
\newcommand{\een}{\end{eqnarray}}
\newcommand{\be}{\begin{equation}}
\newcommand{\ee}{\end{equation}}
\newcommand{\ba}{\begin{eqnarray}}
\newcommand{\ea}{\end{eqnarray}}

\begin{document}

\title{Wormholes respecting energy conditions and solitonic shells in DGP gravity}


\author{Mart\'{\i}n G. Richarte}\email{martin@df.uba.ar}
\address{ Departamento de F\'{\i}sica, Facultad de Ciencias Exactas y
Naturales,  Universidad de Buenos Aires, Ciudad
Universitaria, Pabell\'on I, 1428 Buenos Aires, Argentina}

\begin{abstract}
We build spherically symmetric wormholes within the DGP theory. We calculate the energy localized on the shell, and we find that for certain values of the parameters wormholes could be supported by matter not violating the energy conditions. We also show that it could exist solitonic  shells charaterized by zero pressure and zero energy; thereafter  we make some  observations regarding their  dynamic on the  phase plane. 
\end{abstract}

\maketitle

\section{Introduction}
 Traversable Lorentzian wormholes  \cite{motho,visser} are topologically non trivial solutions of the equations of gravity which would imply a  connection between two regions of the same universe, or of two universes, by a traversable  throat.  In the case that such geometries actually exist they could show some interesting peculiarities as, for example, the possibility of using them for  time travel \cite{morris,novikov}. A basic difficulty with wormholes is that the  flare-out condition \cite{hovis1} to be  satisfied at the throat  requires the presence of matter which violates the energy conditions (``exotic matter'') \cite{motho,visser,hovis1,hovis2}. It was recently shown \cite{viskardad}, however, that the amount of exotic matter necessary for supporting a wormhole geometry can be made infinitesimally small. Thus, in subsequent works special attention has been devoted to quantifying the amount of exotic matter  \cite{bavis,nandi1}, and  this measure of the exoticity has been pointed as an indicator of the physical viability of a traversable wormhole \cite{nandi2}.

A central aspect of any solution of the equations of gravitation is its mechanical stability. The stability of wormholes  has been thoroughly studied for the case of small perturbations preserving the original symmetry of the configurations. In particular, Poisson and Visser \cite{poisson} developed a straightforward approach for analyzing this aspect for thin-shell wormholes, that is, those which are mathematically constructed by cutting and pasting two manifolds to obtain a new manifold \cite{mvis}. In these wormholes the associated  supporting matter is located on a shell placed at the joining surface; so the theoretical tools for treating them is the Darmois--Israel formalism, which leads to the Lanczos equations \cite{daris,mus}. The solution of the Lanczos equations gives the dynamical evolution of the wormhole once an equation of state for the matter on the shell is provided.  Such a procedure has been subsequently followed  to study the stability of more general spherically  symmetric configurations (see, for example, Refs. \cite{eirom}).

Wormholes in theories beyond Einstein framework  have gained a lot of interest in the last years  because they  seem to possess some curious  properties regarding the kind of matter which could support them. A few  examples of these alternatives theories  are the Einstein--Gauss-Bonnet picture \cite{mi1, whegb1, whegb2}, scalar-tensor theories \cite{mi2,whbdicke}, $F(R)$-theory or massive gravity \cite{whfrotros}. In particular, for the Einstein--Gauss--Bonnet theory, it was shown that static thin-shell wormholes could be supported by ordinary matter respecting the energy conditions\cite{mi1}. Moroever, $C^2$-type womholes with the latter property can also exist once the nonlinear Gauss-Bonnet term is included in the field equations \cite{whegb1}. Of course, this feature is not only exclusive of the Gauss-Bonnet paradigms; being the Brans-Dicke gravity  another set up where the thin-shell wormholes fulfil weak and null  energy conditions \cite{mi2}. 

In addition,  a new type of gravitational model was widely studied in the context of cosmology as well as particle physics, the so called Dvali, Gabadadze and Porrati (DGP) theory. It predicts deviations from the standard 4D gravity over large distances. The transition between four and higher-dimensional gravitational potentials in the
DGP model arises because of the presence of both the brane and bulk Hilbert--Einstein (H--E) terms in the action \cite{dgp}.
 Cosmological considerations of the DGP model were first discussed in \cite{ace1,ace2} where it was
shown that in a Minkowski bulk spacetime we can obtain self-accelerating solutions. In the original
DGP model it is known that 4D general relativity is not recovered at linearized level. However, some
authors have shown that at short distances we can recover the 4D general relativity in a spherically
symmetric configuration (see for example \cite{tanaka}).

It is worth mentioning that an interesting feature of the original DGP model is the existence of
ghost-like excitations \cite{exic,jc1}. Further,  the viability of the self-accelerating cosmological solution  in the DGP gravity was carefully studied in \cite{dimitri}. For a comprehensive review of the existence of 4D ghosts on the self-accelerating branch of solutions in DGP models see \cite{jc2}.

A common feature amongs alternative theories is that the junction conditions for the thin-shell wormholes  are modified considerably, adding new types of geometrical objects besides the usual extrinsic curvature. The contributions form the curvature tensors, theoretically, seem to allow the existence of  wormholes supported by ordinary matter. For all these reasons, we consider that  the construction of  wormholes within DGP gravity deserves to be examined in detail  to conclude whether or not they could  fulfill the energy conditions.

In this work we explore thin-shell wormholes within the DGP gravity theory. Our research is focus on configurations supporting by non-exotic matter which satisfy the energy conditions. Then, we show the existence of solitonic vacuum shells  and  make some comment about their dynamic.

\section{Five-dimensional bulk solution}
We start from the action for the DGP theory in five-dimensional manifold ${\cal{M}}_5$  with four-dimensional boundary $\partial{\cal{M}}_5=\Sigma$ (cf. \cite{jc1}),
\begin{eqnarray*}
\label{dgpa}
S&=&2M^{3}_{5}\int_{{\cal M}_{5}} d^{5}{x}\sqrt{-g}R(g_{\mu\nu}) + \int_{\Sigma} d^{4}{x}\sqrt{-\gamma}2M^{2}_{4}{\cal R}(\gamma_{ab}) \\ 
&+&\int_{\Sigma} d^{4}{x}\sqrt{-\gamma}\Big(-4M^{3}_{5}{\cal{K}}(\gamma_{ab}) + {\cal{L}}_{m}\Big),\\
\end{eqnarray*}
where $g_{\mu\nu}$ is the five-dimensional metric, $\gamma_{ab}$ is the four-dimensional induced metric on the boundary $\Sigma$, and $\cal{K}$ is the trace of extrinsic curvature. 
The extra term in the boundary introduces a mass scale $m_{c}=2M^{3}_{5}/M^{2}_{4}=r^{-1}_{c}$, that is, the model has one adjustable parameter, namely $m_c$ which determines a scale that separates two different regimes of the theory. For distances much smaller than $m^{-1}_{c}$  one would expect the solutions to be well approximated by General Relativity and the modifications to appear at larger distances. This is indeed the case for distributions of matter and radiation which are homogeneous and isotropic
at scales $\gtrsim r_{c}$. Tipically, $m_{c}\sim 10.42 \mbox{GeV}$, so it sets the distance/time scale $r_{c}=m^{-1}_{c}$ at which the Newtonian potential significantly deviates from the conventional one \footnote[1]{However, a  more careful analysis indicates that when  compact static source of the mass M and radius $r_{0}$ are taken into account, such that
$r_{M} < r_{0} << r_{c}$ ($r_M= 2MG_N$ is the Schwarzschild radius) a new scale, combination
of $r_c$ and $r_M$, emerges (the so-called Vainshtein scale) : $r_{*}=(r^{2}_{c}r_{M})^{1/3}$.
Below this scale the predictions of the theory are in good agreement with the GR
results and above it they deviate considerably(cf.\cite{foot1})}.

It is a well known fact that the DGP scheme is a five-dimensional model where gravity propagates throughout an infinite bulk, and matter fields in ${\cal{L}}_{m}$ are confined to a 4-dimensional boundary. The action for gravity at lowest order in the derivate expansion is a bulk Einstein-Hilbert term and a boundary one, generically with two different planck masses $M_5$, $M_4$, plus a suitable Gibbons-Hawking term.
In the bulk the DGP  equations are the Einstein ones in vacuum: $G^{(5)}_{\mu\nu}=0$. Then in this case, the Birkhoff's theorem forces the bulk metric to be static, and of the  Schwarzschild form:
\begin{eqnarray}
\label{metric1} 
ds^{2}&=& -f(r) dt^{2} + \frac{1}{f(r)}dr^{2} + r^{2}d\Omega^{2}_{3},
\\
\label{metric2} 
f(r)&=&1- \frac{\mu}{r^2}
\end{eqnarray}
where the parameter $\mu$ is related to the five dimensional Arnowitt--Deser--Misner (ADM) mass, $M_{_{ADM}}=3\pi^2\mu M^3_{5}$. The above spacetime has only one horizon placed at $r_{+}=\sqrt{\mu}$ with $\mu>0$. Besides, when $\mu<0$ the manifold only presents a naked singularity at the origin $r=0$.

\section{Wormholes in DGP}
\subsection{Thin-shell construction}
Employing the metric Eqs.(\ref{metric1}-\ref{metric2}) we build a spherically thin-shell wormhole in DGP theory. we take two copies of the spacetime amd remove from each manifold the five-dimensional regions described by 
\begin{equation}
{\cal M}_{\pm}=\left\{x/r_{\pm}\leq a,a>r_{h}\right\}.
\end{equation} 
The resulting manifolds have boundaries given by the timelike hypersurfaces 
\begin{equation}
\Sigma_{\pm}=\left\{x/r_{\pm} = a,a>r_{h}\right\}.
\end{equation}
Then we identify these two timelike hypersurfaces to obtain a geodesically complete new manifold ${\cal {M}}={\cal {M}}^{+}\cup {\cal {M}}^{-}$. We take values of $a$ large enough to avoid the presence of singularities and horizons in the case that the geometry (\ref{metric2}) has any of them. The manifold $\cal M$ repesents a wormhole  with a throat placed at the surface  $r=a$, where the matter supporting the configuration is located. This manifold is constituted by two regions which are  asymptotically flat (see Fig. 1).
\begin{figure}[!h]
\begin{center}
\includegraphics[height=9cm, width=8cm]{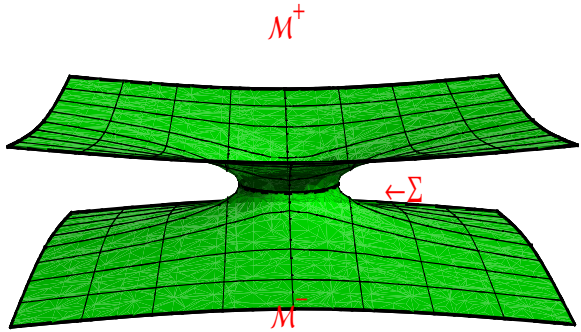}
\caption{We show the wormhole geometry obtained after performing the cut and paste procedure. The shell on $\Sigma$ is located at the throat radius $r=a$.}
\end{center}
\end{figure}
The wormholes throat $\Sigma$ is a synchronous timelike hypersurface, where we define locally a chart with coordinates $\xi^{a}=(\tau, ,\chi,\theta,\phi)$, with $\tau$ the proper time on the shell.  Though  we shall first focus in static configurations, in the susbsequent  we could allow the radius of the throat be a function of the proper time for studying the dynamics evolution of the wormholes, then in general we have that the  boundary hypersurface reads:
\begin{equation}
\Sigma: {\cal H}(r,\tau)=r-a(\tau)=0.
\end{equation}
 It is important to remark that the geometry remains static outside the throat, regardless the radius $a(\tau)$ can vary with time, so no gravitational waves are present.This is naturally guaranteed because the Birkhoff theorem holds for the original manifold.

Our starting point is to list the main geometric objects which shall appear in the junction condition associated with the field equation for $\Sigma$.
The extrinsic curvature,namely ${\cal K}_{ab}$, associated with the two sides of the shell are defined as follows:

\begin{equation}
{{\cal K}}^{\pm}_{ab}=-n^{\pm}_{\kappa}\left(\frac{\partial^{2}X^{\kappa}}{\partial\xi^{a}\partial\xi^{b}}+\Gamma^{\kappa}_{\mu\nu}\frac{\partial X^{\mu}}{\partial\xi^{a}}\frac{\partial X^{\nu}}{\partial\xi^{b}}\right)_{r=a},
\end{equation} 
where $n^{\pm}_{\kappa}$ are the unit normals ($n_{\kappa}n^{\kappa}=1$) to the surface $\Sigma$ in ${\cal M}$:
\begin{equation} 
n^{\pm}_{\kappa}=\pm\left| g^{\mu\nu}\frac{\partial {\cal H}}{\partial X^{\mu}}\frac{\partial {\cal H}}{\partial X^{\nu}} \right|\frac{\partial {\cal H}}{\partial X^{\kappa}}
\end{equation}

The  field equations projected on the shell $\Sigma$ are the generalized junction (or Darmois--Israel) conditions \cite{jc1,jc2} 
\begin{equation}
\label{jc}
r_{c}\Big({\cal R}_{ab}-\frac{1}{2}\gamma_{ab}{\cal R}\Big)-2\Big(\left\langle {\cal K}_{ab}-{\cal K}\gamma_{ab}\right\rangle\Big)=\frac{{\cal S}_{ab}}{8M^3_{5}},
\end{equation}
where the bracket $\left\langle .\right\rangle$ stands for  the jump of a given quantity across the  hypersurface $\Sigma$ and  $\gamma_{ab}$ is the induced metric on $\Sigma$.
Notice that the first term in (\ref{jc}) is not enclosed with the brackets because this contribution comes from the  four dimensional E-H term in the DGP action (\ref{dgpa}) which already lives in the boundary so it does not need to be projected on $\Sigma$. By taking the limit  $r_{c}\rightarrow 0$ we recover the standard Darmois--Israel junction condition found in \cite{daris}.

Now, let us calculate some quantities that we shall need later. The mixed components of the four-dimensional Einstein tensor are given by
\begin{eqnarray}
\label{ET} 
{\cal G}^{0}_{~0}&=&-3\Big(\frac{{\dot{a}}^{2}}{a^2} + \frac{1}{a^2}\Big),
\\
\label{EE} 
{\cal G}^{i}_{~j}&=&-\Big(\frac{1}{a^2}+ \frac{{\dot{a}}^{2}}{a^2}+ 2\frac{\ddot{a}}{a}\Big)\delta^{i}_{~j}
\end{eqnarray}
 where dot means derivate with respect to the proper time on $\Sigma$. The extrinsic curvature components read
\begin{eqnarray}
\label{KT} 
\left\langle {\cal K}^{0}_{~0} \right\rangle&=&\frac{2\ddot{a} + f'(a)}{\sqrt{f(a)+{\dot{a}}^{2}}},
\\
\label{KE} 
\left\langle {\cal K}^{i}_{~j}\right\rangle&=&\frac{2}{a}\sqrt{f(a)+{\dot{a}}^{2}}~\delta^{i}_{~j}
\end{eqnarray}
 where the prime inidcates the derivates with respect to $a$. The  most general form of the stress energy tensor on shell compatible with the simetries is 
 \begin{equation}
 \label{tem}
 {\cal S}^{a}_{~b}=~\mbox{diag}~(-\sigma, p ~\delta^{i}_{~j})
\end{equation}
where $\sigma$ is the energy density and $p$ is the pressure. Replacing Eqs(\ref{EE}-\ref{tem}) into the DGP junction condition(\ref{jc}) we obtain that the energy density and the pressure can be recast as
\begin{eqnarray}
\label{sigma} 
\frac{\sigma}{8M^3_{5}}&=&3r_{c}\Big(\frac{{\dot{a}}^{2}}{a^2} + \frac{1}{a^2}\Big)-\frac{12}{a}\sqrt{f(a)+{\dot{a}}^{2}},
\\
\label{pe} 
\frac{p}{8M^3_{5}}&=&-r_{c}\Big(\frac{{\dot{a}}^{2}}{a^2} + \frac{1}{a^2}+ \frac{2\ddot{a}}{a}\Big)+\frac{8}{a}\sqrt{f(a)+{\dot{a}}^{2}}\\ 
&+&2\frac{2\ddot{a}+f'}{\sqrt{f(a)+{\dot{a}}^{2}}}.
\end{eqnarray}
where the DGP contributions are encoded in the $r_{c}$ factor of the above equations. If we take  $r_{c}\rightarrow 0$ in both equations (\ref{sigma}) and (\ref{pe}) we recover  the expression for the energy density $\sigma$ and the pressure $p$ found in \cite{mi1}; ignoring the Gauss-Bonnet contribution.

In order to carry on let us comment that we still have the  usual energy conservation,$\nabla_{a}{\cal S}^{ab}=0$ by virtue of $\nabla^{a}({\cal K}_{ab}-\gamma_{ab}{\cal K})=0$, comig from the momentum constraint implicit in the five-dimensional Einstein equations. Further it is easy to see from $\sigma$  and $p$ that the energy conservation equation is fulfilled:
\begin{equation}
\frac{d(a^{3}\sigma)}{d\tau}+p\frac{da^{3}}{d\tau}=0,\label{conserva}
\end{equation}
the first term in Eq. (\ref{conserva}) represents the
internal energy change of the shell and the second the work by internal forces of the
shell. The dynamical evolution of the wormhole throat is governed by the generalized Lanczos equations and to close the system we must supply an equation of state $p = p(\sigma)$ that relates $p$ and $\sigma$. Notice that the reason why one obtains exact conservation, i.e, no energy flow to the bulk, is that the normal-tangential components of the stress tensor in the bulk is the same on both side of the junction hypersurface.

\section{Matter supporting the wormholes}
Recently, classical solutions  within  the DGP model were found when the stress-energy tensor on the brane satisfies the dominant energy condtion, yet the brane has negative energy from the bulk point of view (see \cite{jc1}). Within this frame, the study of superluminal propagation  indicates  that superlumnosity occurs whenever the stress tensor on the shell is a pure cosmological constant, irrespective of the value  of the shell density (cf.\cite{jc1}). All these elements are  good reasons to consider a careful discussion about the nature of matter supporting wormholes in the DGP model.
Moreover, motivated by the results within Einstein-Gauss–Bonnet gravity (i.e. with $R^{2}$-like terms) in \cite{mi2},
here we evaluate the amount of exotic matter and the energy conditions, following the approach
presented above where the four-dimensional H--E term generalizes the standard junction, adding  a few  geometrical terms, which indeed represents  the Einstein tensor projected on the shell.  Consequently, coming the DGP contribution from the curvature tensor, the next approach is clearly the most suitable to give a precise meaning to the characterization of matter supporting the wormhole.

The \emph{weak energy condition} (WEC) states that for any timelike vector $U^{\xi}$ it must be $T_{\xi\eta}U^{\xi}U^{\eta}\geq 0$;
the WEC also implies, by continuity, the \emph{null energy condition} (NEC), which means that for any null
vector $k^{\xi}$ it must be $T_{\xi\eta}k^{\xi}k^{\eta}\geq 0$ [3]. In an orthonormal basis the WEC reads $\rho\geq 0$, $\rho + p_{l}\geq 0$ $\forall ~ l$  while the NEC takes the form $\rho + p_{l}\geq 0$ $\forall ~ l$.  Besides, the \emph{strong energy condition} states that $\rho + p_{l}\geq 0$ $\forall ~ l$, and $\rho + 3p_{l}\geq 0$ $\forall ~ l$.

In the case of thin-shell wormholes the radial pressure $p_r$ is zero,  within Einstein gravity, the surface energy density must fulfill $\sigma < 0$, so that both energy conditions would be violated. The sign of $\sigma+p_{t}$ where $p_t$ is the transverse pressure is not fixed, but it depends
on the values of the parameters of the system. In what follows we restrict to static configurations. The surface energy density $\sigma_{0}$ and the
transverse pressure $p_{0}$ for a static configuration ($a = a_0$, $\dot{a}=0$, $\ddot{a}=0$) are given by

\begin{eqnarray}
\label{sigo} 
\frac{\sigma_{0}}{8M^3_{5}}&=& \frac{3r_{c}}{a^{2}_{0}}-\frac{12}{a_{0}}\sqrt{f(a_{0})},
\\
\label{po} 
\frac{p_{0}}{8M^3_{5}}&=&- \frac{r_{c}}{a^{2}_{0}}+\frac{8}{a_{0}}\sqrt{f(a_{0})}+ 2\frac{f'(a_{0})}{\sqrt{f(a_{0})}}.
\end{eqnarray}
Now the sign of the surface energy density  as well as the pressure is, in principle, not fixed. The most usual choice  for quantifying the  amount of exotic matter in a Lorentzian wormhole is the integral \cite{nandi1}:
\begin{equation}
\Omega= \int (\rho + p_{r})\sqrt{-g_{5}}\,d^{4}x.
\end{equation}
We can introduce a new radial coordinate $R=\pm(r-a_{0})$ with $\pm$ corresponding to each side of the shell. Then,  
because  in our construction the energy density is located on the surface, we can also write $\rho=\delta(R)\,\sigma_{0}$, and because the shell does not exert radial pressure  the amount of exotic matter reads
\begin{equation}
\Omega=\int\limits^{2\pi}_{0} \int\limits^{\pi}_{0}\int\limits^{\pi}_{0}\int\limits^{+\infty}_{-\infty}\delta(R)\,\sigma_{0} \sqrt{-g_{5}}\, d R\, d\xi\,d\theta\,d\phi\ =2\pi^{2}  a_{0}^3 \sigma_{0}.
\end{equation}
Replacing the explicit form of $\sigma_{0}$ and $g_{5}$, we obtain the exotic matter amount as a function of the parameters  that characterize the configurations:
\begin{equation}
\label{ome1}
\Omega=16M^3_{5}\pi^{2}\Big(~3r_{c}~ a_{0}-12a^{2}_{0}\sqrt{ f(a_{0})}~\Big).
\end{equation}
where $f$ is  given by the bulk solution. For $r_{c}\rightarrow 0$ we obtain the exotic amount for Schwarzschild geometries  as if it was calculated  with the standard junction conditions.
Far away from the General Relativity limit  we now find that there exist positive contributions to  $\sigma_{0}$; these come from the different
signs in the expression (\ref{ome1}) for the surface energy density, because is proportional to $\sigma_{0}$. We stress
that this would not be possible if the standard Darmois–-Israel formalism was applied, treating the
DGP contribution as an effective energy-momentum tensor, because this leads to $\sigma_{0}\propto - \sqrt{ f(a_{0})}/a_{0}$. Now, once the explicit form of the function $f(a_{0})$ is introduced in Eq.($\ref{ome1}$), we focus on  what are the conditions  that lead to wormholes with $\sigma_{0}>0$ or $\Omega>0$. Then, it can be proved that wormholes with a non-negative surface density located at the shell are  allowable when the following inequalities are simultaneously satisfy: 

\begin{eqnarray}
\label{snula2} 
\frac{r_{c}}{a^{2}_{0}}-\frac{4}{a_{0}}\Big(1-\frac{\mu}{a^{2}_{0}}~\Big)^{1/2}&>&0,
\\
\label{fueho1} 
a^2_{0}-\mu&>0&,
\end{eqnarray}
so it is always possible  to choose $a_{0}$ such that  the existence  of thin-shell wormholes is compatible with positive  surface energy desnsity (see Fig2.), more precisely its radius  must belong to  the interval given below:
\begin{equation} 
 \label{int1}
 \sqrt{\mu}<a_{0}\leq \big(\mu + \frac{r^{2}_{c}}{16}\big)^{\frac{1}{2}}
\end{equation}
Notice that $r_{c}$-term is essential to have positive energy density; as one would expect, in the limit $r_{c} \longrightarrow 0$, this possibility completly vanishes. 
\begin{figure}[!h]
\includegraphics[height=8cm, width=7cm]{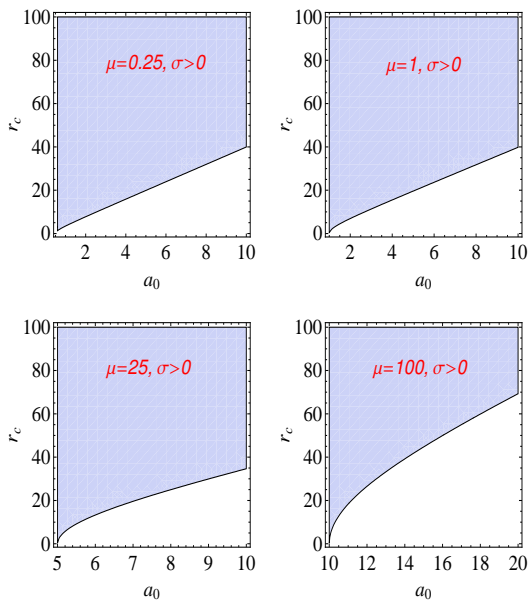}
\caption{We plot the zones in the plane $r_{c}-a_{0}$ where the condition $\sigma_{0}>0$ for several values of $\mu$.}
\end{figure}
Besides, from Eq.(\ref{sigo}) and Eq.(\ref{po}) we have that  the sum of the pressure and energy density  takes the form
\begin{equation} 
 \label{int1}
 \sigma_{0}+ p_{0}=8M^3_{5}\left(\frac{2r_{c}}{a^{2}_{0}}+\frac{2a_{0} f'(a_{0})-4f(a_{0})}{a_{0}\sqrt{f(a_{0})}}\right)
\end{equation}
because the first  term in (\ref{int1}) is positive the sign of $\sigma_{0}+ p_{0}$ depends on the second term, implying that the sum is positive  for $\sqrt{\mu}<a_{0}\leq \sqrt{2\mu}$. Therefore, the remarkable result is that we have a region with $\sigma_{0}\geq0$ and besides $\sigma_{0}+ p_{0}\geq0$ , so the WEC and the NEC are satisfied (see Fig.3 and Fig.4). Additionally, it is easy to corroborate that $\sigma_{0}+ 3p_{0}=12\times8M^3_{5}/(a_{0}\sqrt{f(a_{0})}) $, then  SEC holds in the interval $a_{0} \in (\sqrt{\mu}, \sqrt{2\mu}]$ (see Fig.3 and Fig.4). Thus, by treating the DGP contribution as a geometric object, the generalized junction conditions (\ref{jc}) provide  a clear meaning to the matter in the shell leading to a central finding that in the  DGP gravity the violation of the energy conditions could be avoided and wormholes could be supported by ordinary matter.

\begin{figure}[!h]
\includegraphics[height=7cm, width=9cm]{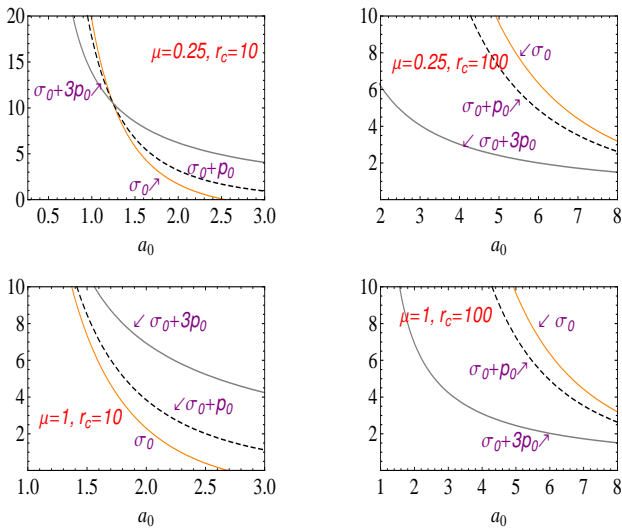}
\caption{We show  $\sigma_{0}$, $\sigma_{0}+p_{0}$ and $\sigma_{0}+3p_{0}$ versus the wormhole radius $a_{0}$ for several values of $(\mu,r_{c})$.}
\end{figure}

\begin{figure}[!h]
\includegraphics[height=7cm, width=9cm]{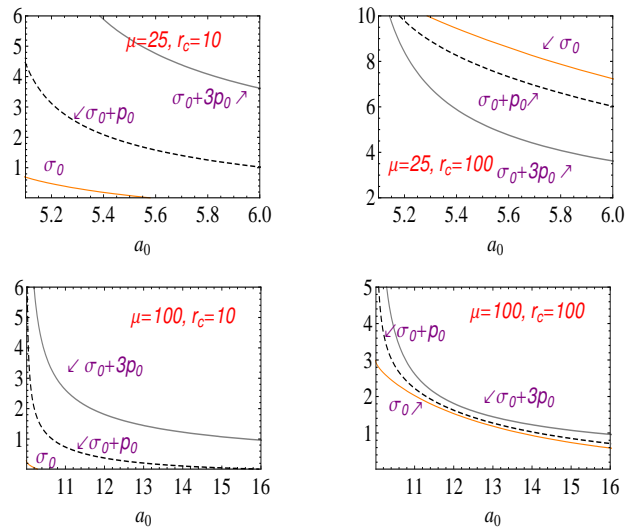}
\caption{We show  $\sigma_{0}$, $\sigma_{0}+p_{0}$ and $\sigma_{0}+3p_{0}$ versus the wormhole radius $a_{0}$ for different values of $(\mu,r_{c})$.}
\end{figure}
However, note that one could choose another route beacuse  Eq.(\ref{jc}) can be formally recast as follows 
\begin{eqnarray}
-16M^3_{5}\left\langle {\cal K}_{ab}-{\cal K}\gamma_{ab}\right\rangle&=&{\cal S}^{eff}_{ab},
\\
{\cal S}_{ab}-8M^3_{5}r_{c}\Big({\cal R}_{ab}-\frac{1}{2}\gamma_{ab}{\cal R}\Big)&=&{\cal S}^{eff}_{ab}
\end{eqnarray}
although this identification is also possible; physically we would be  treating curvature objects  as an effective source for the junction condition. Moreover, based on effective energy-momentum tensor approach we  inevitably would obtain that the energy density is  negative definite because the flare-out condition is fulfilled. For a review  of junction conditions within the DGP theory see \cite{jc3} and references therein.

\section{Solitonic wormholes/shell}
In general to obtain the dynamic picture  of the wormholes within the DGP gravity is a very complicated task. As it  can see from the Eqs. (\ref{sigma}-\ref{pe}) nonlinear character of these expressions make  the standard procedure exposed in \cite{ersswh} very hard to implement. So, we are going to focus in a particular type of wormholes/shell. To be precise we desire to eximane if it is possible to have dynamical solitonic wormholes/shells characterized by a zero pressure ($p=0$) and zero energy density ($\sigma=0$).
Unlike the standard Darmois--Israel junction condition, nontrivial  solutions may be possible even when ${\cal S}^{a}_{b}=0$. That is, the extrinsic curvature can be discontinuos across the throat with no matter on the shell to serve a source; turning the discontinuity a self-supported gravitational system. Of course, these configurations are impossible in the Einstein gravity but not in the Einstein-Gauss-Bonnet gravity (cf. \cite {whegb2}).

For $\dot{a}\neq 0$ the Eq.(\ref{conserva}) shows that  if $\sigma=0$ then $p=0$; so we are going to work with the most useful expression which in this case is  given by $\sigma$.  Following the procedure  mentioned in \cite{jc2}  we shall plot  trayectories in the  phase space spanned by $(\dot{a},a)$. Because of the energy constraint ($\sigma=0$) is invariant under the symmetry $\dot{a}\longleftrightarrow -\dot{a}$  we can work on a two-dimensional plane which is defined as a non-compact domain, namely ${\cal B}=(0, +\infty) \times (r_{+}, +\infty)$. The curves which represent  the dynamics of solitonic wormholes are obtained by imposing the following conditions:
\begin{eqnarray}
\label{snula} 
r_{c}\big({\dot{a}}^{2} + 1\big)-4\Big(a^2-\mu +a^2{\dot{a}}^{2}\Big)^{\frac{1}{2}}&=&0,
\\
\label{fuerahorionte} 
a^2-\mu&>0&
\end{eqnarray}
such that the first inequality guarantees zero energy  whereas the second one ensures that the wormholes radius is larger than event horizon.
\begin{figure}[!h]
\label{sp}
\includegraphics[height=11cm, width=9cm]{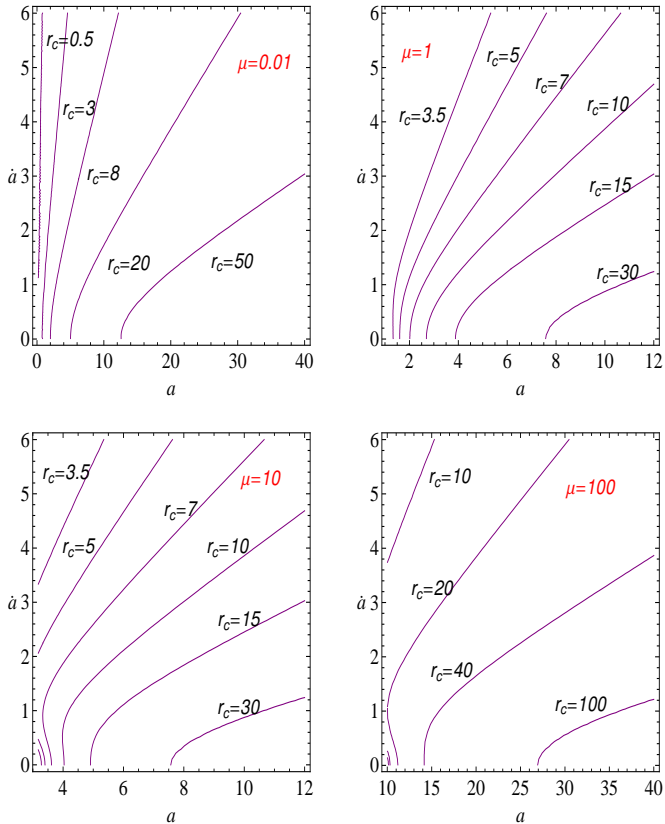}
\caption{We show the trayectories of vacuum shells in the fase plane for $\mu \in [0.01,100]$ and different values of DGP scale $r_{c}$.}
\end{figure}
According to  Fig. 5, the  phase diagrams shows that for small or large $\mu$ and with a DGP scale covering the interval $[0.5,50]$, the shell velocity is a monotone increasing function for $\dot{a}\in(0,+\infty)$ (or decreasing one  when $\dot{a} \in (-\infty, 0)$). Notice that the same conclusion is obtained when  the parameter $r_{c}$ takes larger values. In order to see if these types of shells speed up or decelerate we use the zero pressure condition to get a functional relation $\ddot{a}={\cal N}[a,\dot{a}]$ which determines the sign of $\ddot{a}$:
\begin{equation}
\label{signAc}
{\cal N}=\frac{-8af(a)-8a{\dot{a}}^{2}-2a^{2}f'(a)+r_{c}(1+{\dot{a}}^{2})\sqrt{f(a)+{\dot{a}}^{2}}}{2a\Big(2a- r_{c}\sqrt{f(a)+{\dot{a}}^{2}}\Big)}
\end{equation}
For all $\mu$ and $ r_{c}$ considered in this section  we obtain that the kinematic of the shell has four possible types of dynamical evolution. More precisely, the solitonic solution  could suffer an accelerated ($\ddot{a}>0$) or decelerated ($\ddot{a}<0$) expansion ($\dot{a}>0$)  as well as  an accelerated or decelerated contraction ($\dot{a}<0$) regimes.

Unlike the Einstein--Gauss-Bonnet case  studied in \cite{whegb2} it turns  that the existence of solitonic shells in  DGP gravity  does not require the presence of a cosmological constant term in the bulk spacetime.  

\section{Summary}
The generalization of Einstein gravity in the way proposed by  Dvali, Gabadadze and Porrati (DGP)  introduces a new parameter, which allows for more freedom in the framework of determining the most viable  wormhole configurations. If  wormholes could actually exist, one would be interested in those which are  require as little amount of exotic matter as possible. Of course, the case could be that a given change of the theory leads to a worse situation, i. e. that configurations  require more matter violating the energy conditions as the departure from the standard theory becomes relevant. 
However, for suitable wormhole radius, this  seems not to be the case with DGP gravity : Here we have examined the ``exotic'' matter content of thin-shell wormholes using the generalized junction condition, and we  have found that for large values of the DGP parameter, corresponding to a situation far away from the General Relativity limit, the amount of exotic matter is reduced in relation with the standard case beause it can be positive definite. Moreover, the remarkable result is that we have a region with $\sigma_{0}\geq0$ and besides $\sigma_{0}+ p_{0}\geq0$ , so the WEC and the NEC are satisfied. Further the SEC condition holds also. Thus if the requirement of exotic matter is considered as the hardest objection against wormholes, our results suggest that in a physical scenario with small crossover scale ($r_{c}\sim O(1)$) or far away from the General Relativity limit where the DGP becomes dominant  ($r_{c}\gtrsim~10^{2}$) these types of wormholes could be  possible.    
Finally, we showed the existence of  gravitational solitonic wormholes/shell characterized by $\sigma=p=0$  within the DGP model. Unlike the case of Einstein--Gauss-Bonnet theory we found that the existence of  solitonic shells in  DGP gravity  does not  require the presence of a cosmological constant term in the bulk solution.  

\acknowledgments
MGR thanks the University of Buenos Aires for partial support under
project X044.  MGR is also supported by  the Consejo Nacional de Investigaciones Cient\'{\i}ficas y T\'ecnicas (CONICET). 


\end{document}